# Ground state properties of one-dimensional and two-dimensional Hubbard model from Gutzwiller conjugate gradient minimization theory

*Zhuo Ye,[1,*] Feng Zhang,[1] Yong-Xin Yao,[1,*] Cai-Zhuang Wang,[1] Kai-Ming Ho[1]*

[1] Ames Laboratory – US DOE and Department of Physics and Astronomy, Iowa State University, Ames, Iowa 50011, United States

*E-mail: zye@iastate.edu (Z Ye), ykent@iastate.edu (YX Yao)

Abstract

We introduce Gutzwiller conjugate gradient minimization (GCGM) theory, an *ab initio* quantum many-body theory for computing the ground-state properties of infinite systems. GCGM uses the Gutzwiller wave function but does not use the commonly adopted Gutzwiller approximation (GA), which is a major source of inaccuracy. Instead, the theory uses an approximation that is based on the occupation probability of the on-site configurations, rather than approximations that decouple the site-site correlations as used in the GA. We test the theory in the one-dimensional and two-dimensional Hubbard models at various electron densities and find that GCGM reproduces energies and double occupancies in reasonable agreement with benchmark data at a very small computational cost.

1. INTRODUCTION

Strongly correlated materials have attracted increasing interest as they exhibit extraordinary and fascinating electric and magnetic properties. However, the fundamental feature that defines these materials cannot be described with non-interacting entities. Accurate description of electronic correlation effect is required, and thus often impose a formidable computational burden. We cannot place our hope on the fast growth of computer technology. Solving the Schrodinger equation accurately for bigger systems is "a catastrophe of dimension"[1] even for modern computers. Study of strongly correlated materials calls for development of approaches that are both fast and reasonably accurate to describe the electronic correlation in real materials.

Moving towards this goal, we have been developing numerical tools in recent years based on Gutzwiller variational wave function (GWF) that were proposed by Gutzwiller five decades ago.[2,3,4] The GWF introduces correlations into a trial non-interacting wave function via an onsite correlation factor. Since a closed form of expectation value of an observable with respect to GWF is not generally available, Gutzwiller then introduced a further approximation called "Gutzwiller approximation" (GA) to estimate the energy. Some of our earlier work includes the development of hybrid approaches that combine density functional theory (DFT) and GA (DFT+G, see Ref. [5,6,7,8,9] for work by our group as well as other researchers). These approaches had been demonstrated to be very effective to describe the strong electronic correlation. However, DFT+G approaches and other hybrid approaches, that integrate DFT with many-body techniques

[10,11,12,13,14], often contain adjustable screened Coulomb parameters, which limits the predictive power. A challenging goal in studies of strongly correlated materials is understanding the fundamental science of existing strongly correlated materials. A more difficult goal lies beyond understanding and is to predict novel properties of new materials and structures, which calls for *ab initio* methods without any adjustable parameters. Speaking of *ab initio* methods, one needs to particularly consider the computational cost because they often take larger amounts of computer time, memory, and disk space. We chose GWF since the GWF and GA-based schemes had proven to be very computationally efficient. Nevertheless, after a careful examination with GA, we found that GA is a major source of inaccuracy. In GA, an element of one-particle density matrix expressed in the GWF can be approximated by the multiplication of the corresponding element in the trial non-interacting wave function and the one-electron renormalization $z$-factor that is assigned to each of the two relevant sites. In this way, the approximation "decouples" the two correlated sites by using a site-site factorization. This inter-site decoupling approximation lacks the flexibility to be further generalized for better accuracy, because it overlooks the site-site correlation, given that GWF already has the same problem by introducing correlations into a non-interacting wave function only via an onsite correlator.

That was the motivation that we developed most recently the Gutzwiller conjugate gradient minimization (GCGM) theory. We still used the GWF, but goes beyond the GA. Since a closed form of expectation value with respect to GWF is not generally available, and exact numerical evaluations require the variational Quantum Monte

Carlo (QMC) simulations that can be quite time-consuming, new approximations need to be made to provide efficient and reasonably accurate evaluations. Within GCGM, we made such approximations and demonstrated that our algorithm provides improved accuracy by benchmarking it with energy calculations of both ground and excited state, with the focus on small molecules.[15,16,17] We have to point out that for small molecules, GCGM does not have advantages over other quantum chemistry methods. Its computational time is comparable to full configuration interaction (FCI) method. We have focused on molecules just to validate GCGM. We want to emphasize that GCGM has a good scaling with system size[15] and is particularly efficient for a bulk system. For *ab initio* methods, one needs to consider the computational cost when determining whether they are appropriate for a specific problem. For simple system like small molecules, one would choose a method as accurate as possible, like FCI or Quantum Monte Carlo (QMC),[18,19,20] because the power of modern computers allows him to do so. But for complex systems like real bulk materials, a method that is somewhat less accurate but very efficient is more appreciated. In this work, we will demonstrate the potential of GCGM to be such an *ab initio* method that is appropriate for study of bulk systems by testing it in single-band Hubbard models, both 1D and 2D, as the first step[1] of our exploratory towards real bulk materials.

There have been efforts to access Hubbard models using the GWF.[21,22,23,24,25,26] On one hand, the GA has been widely applied although it sometimes fails to give an accurate description of the model system.[21] Some attempts have been made on solving variational problem with the GWF beyond the GA.[25,26] On the other hand, pure GWF, in general, is not a good choice and should be supplemented with some inter-site correlators such as Jastrow correlator.[22] In this work, we propose an approach that goes beyond GA where site-site constraints are applied in the GWF for a better description of Hubbard models. We will show that in most cases, our approach represents a large improvement over the GA.

---

[1] We had presented results of one dimensional hydrogen chain in Ref. [15]. But this is our 1st work to systematically study bulk systems.

2. METHODS

In this section, we introduce to the readers the GCGM approach that we recently developed. The basic ideas, formalism and a thorough discussion of the approach are described along the way. The approach is initially developed for diatomic molecules, where the least approximation is needed. Then the approach is generalized to systems with more than 2 atoms, with approximations introduced to treat the pair-envionment interactions. Some equations had already been presented in our recent works.[15,16,17] Nevertheless, we present all the necessary formulas here to be self-contained.

The full *ab initio* nonrelativistic Hamiltonian for an interacting many-body system can be expressed in the form of second quantization as,

$$\hat{H} = \sum_{i\alpha j\beta,\sigma} t_{i\alpha j\beta} c^\dagger_{i\alpha\sigma} c_{j\beta\sigma} + \frac{1}{2} \sum_{\substack{i\alpha j\beta \\ k\gamma l\delta, \sigma\sigma'}} u(i\alpha j\beta; k\gamma l\delta) c^\dagger_{i\alpha\sigma} c^\dagger_{j\beta\sigma'} c_{l\delta\sigma'} c_{k\gamma\sigma} \qquad (1)$$

where $i, j, k, l$ are the atomic site indices, $\alpha, \beta, \gamma, \delta$ the orbital indices, and

$\sigma, \sigma'$ the spin indices. Here, $t$ and $u$ are the one-electron hopping integral and the two-electron Coulomb integral, respectively. They can be expressed as,

$$t_{i\alpha j\beta} = \left\langle \phi_{i\alpha} \left| \hat{T} + \hat{V}_{ion} \right| \phi_{j\beta} \right\rangle, \qquad (2)$$

$$u(i\alpha j\beta; k\gamma l\delta) = \iint d\mathbf{r} d\mathbf{r}' \phi_{i\alpha}^*(\mathbf{r}) \phi_{j\beta}^*(\mathbf{r}') \hat{U}(\mathbf{r}-\mathbf{r}') \phi_{l\delta}(\mathbf{r}') \phi_{k\gamma}(\mathbf{r}), \qquad (3)$$

where $\hat{T}$, $\hat{V}_{ion}$, and $\hat{U}$ are the operators for kinetic energy, ion-electron interaction and Coulomb interaction, respectively. The Hamiltonian in Eq. (1) does not include any adjustable parameters. In the GCGM approach, the GWF is used throughout the calculation. It is constructed from a noninteracting wave function $|\Psi_0\rangle$, i.e. a single Slater determinant,[2]

$$\left| \Psi_{GWF} \right\rangle = \prod_i \left( \sum_\Gamma g(\Gamma_i) |\Gamma_i\rangle \langle \Gamma_i| \right) |\Psi_0\rangle, \qquad (4)$$

where $g(\Gamma_i)$ is the Gutzwiller variational parameter determining the occupation probability of the on-site configuration $|\Gamma_i\rangle$, defined as a Fock state at $i^{th}$ site $|\Gamma_i\rangle \equiv \prod_{\alpha\sigma \in \Gamma_i} c_{\alpha\sigma}^\dagger |\varnothing\rangle$. Here the creation operator $c_{\alpha\sigma}^\dagger$ creates an electron at the orbital-$\alpha$ with spin-$\sigma$ in the vacuum state $|\varnothing\rangle$. The GWF introduces correlations into the non-interacting wave function via the on-site correlation factor $g(\Gamma_i)$. The total energy without adopting Gutzwiller approximation can be expressed as,

---

[2] $|\Psi_0\rangle$ is a non-interacting wave function (a single Slater determinant) in traditional GWF based methods. However, the current GCGM scheme is more flexible in the selection of $|\Psi_0\rangle$ and does not require it to be a single Slater determinant. For example, one can easily modify $|\Psi_0\rangle$ to include all the configurations of the exact solution when it fails to do so. For more details please check Ref. [15].

$$E_{GWF} = \sum_{i\alpha j\beta,\sigma} t_{i\alpha j\beta} \langle c^\dagger_{i\alpha\sigma} c_{j\beta\sigma} \rangle_{GWF} + \frac{1}{2} \sum_{\substack{ijkl,\alpha\beta\gamma\delta \\ \sigma\sigma'}} u(i\alpha j\beta; k\gamma l\delta) \langle c^\dagger_{i\alpha\sigma} c^\dagger_{j\beta\sigma'} c_{l\delta\sigma'} c_{k\gamma\sigma} \rangle_{GWF}$$

$$\approx \sum_{i\alpha j\beta,\sigma} t_{i\alpha j\beta} \langle c^\dagger_{i\alpha\sigma} c_{j\beta\sigma} \rangle_{GWF} + \frac{1}{2} \sum_{\substack{i,\alpha\beta\gamma\delta \\ \sigma\sigma'}} u(i\alpha i\beta; i\gamma i\delta) \langle c^\dagger_{i\alpha\sigma} c^\dagger_{i\beta\sigma'} c_{i\delta\sigma'} c_{i\gamma\sigma} \rangle_{GWF}$$

$$+ \frac{1}{2} \sum_{\substack{i\alpha j\beta \\ k\gamma l\delta,\sigma\sigma'}}{}' \left( u(i\alpha j\beta; k\gamma l\delta) - \delta_{\sigma\sigma'} u(i\alpha j\beta; l\delta k\gamma) \right) \langle c^\dagger_{i\alpha\sigma} c_{k\gamma\sigma} \rangle_{GWF} \langle c^\dagger_{j\beta\sigma'} c_{l\delta\sigma'} \rangle_{GWF}$$

(5)

Here, for any operator $O$, $\langle O \rangle_{GWF}$ is a short-hand notation for $\langle \Psi_{GWF} | O | \Psi_{GWF} \rangle$. $\sum'$ indicates that the pure on-site terms are excluded from the summation. In Eq. (5), the on-site two particle correlation matrix (2PCM) are treated rigorously and the intersite 2PCM are evaluated using Hartree-Fock(HF)-type factorized approximation (Wick's theorem, see Ref. [27]).

Eq. (5) gives an expression of total energy in terms of one particle density matrix (1PDM) and on-site 2PCM. Next we will start with a diatomic molecule and give expressions of the 1PDM and 2PCM. After that, we will generalize our method to systems of more than 2 atoms and bulk systems. For a system that has only 2 atomic sites, the one-particle density matrix (1PDM) can be expressed as,

$$\langle c^\dagger_{i\alpha\sigma} c_{i\beta\sigma} \rangle_{GWF} = \frac{1}{\langle \Psi_{GWF} | \Psi_{GWF} \rangle} \sum_{\Gamma_i, \Gamma'_i, \Gamma_j} \langle \Gamma_i | c^\dagger_{i\alpha\sigma} c_{i\beta\sigma} | \Gamma'_i \rangle g(\Gamma_i) g(\Gamma'_i) g(\Gamma_j)^2 \xi^0_{\Gamma_i, \Gamma_j, \Gamma'_i, \Gamma_j},$$

(6)

$$\left\langle c_{i\alpha\sigma}^{\dagger} c_{j\beta\sigma} \right\rangle_{GWF} = \frac{1}{\left\langle \Psi_{GWF} | \Psi_{GWF} \right\rangle} \sum_{\Gamma_i, \Gamma_j, \Gamma_i', \Gamma_j'} \left\langle \Gamma_i | c_{i\alpha\sigma}^{\dagger} | \Gamma_i' \right\rangle \left\langle \Gamma_j | c_{j\beta\sigma} | \Gamma_j' \right\rangle \cdot$$
$$g(\Gamma_i) g(\Gamma_j) g(\Gamma_i') g(\Gamma_j') \xi^0_{\Gamma_i, \Gamma_j, \Gamma_i', \Gamma_j'} \quad \text{for } i \neq j \tag{7}$$

where $\xi^0_{\Gamma_i, \Gamma_j, \Gamma_i', \Gamma_j'}$ is predetermined coefficient from $|\Psi_0\rangle$,

$$\xi^0_{\Gamma_i, \Gamma_j, \Gamma_i', \Gamma_j'} = \left\langle \Psi_0 | \Gamma_i, \Gamma_j \right\rangle \left\langle \Gamma_i', \Gamma_j' | \Psi_0 \right\rangle, \tag{8}$$

and

$$\left\langle \Psi_{GWF} | \Psi_{GWF} \right\rangle = \sum_{\Gamma_i, \Gamma_j} \xi^0_{\Gamma_i, \Gamma_j, \Gamma_i, \Gamma_j} g(\Gamma_i)^2 g(\Gamma_j)^2. \tag{9}$$

The on-site 2PCM can be expressed as,

$$\left\langle c_{i\alpha\sigma}^{\dagger} c_{i\beta\sigma'}^{\dagger} c_{i\gamma\sigma'} c_{i\delta\sigma} \right\rangle_{GWF} =$$
$$\frac{1}{\left\langle \Psi_{GWF} | \Psi_{GWF} \right\rangle} \sum_{\Gamma_i, \Gamma_i', \Gamma_j} \left\langle \Gamma_i | c_{i\alpha\sigma}^{\dagger} c_{i\beta\sigma'}^{\dagger} c_{i\gamma\sigma'} c_{i\delta\sigma} | \Gamma_i' \right\rangle g(\Gamma_i) g(\Gamma_i') g(\Gamma_j)^2 \xi^0_{\Gamma_i, \Gamma_j, \Gamma_i', \Gamma_j}. \tag{10}$$

Substituting Eqs. (6),(7),(10) into Eq. (5), $E_{GWF}$ can be expressed explicitly as a function of $\{g(\Gamma_i)\}$. $E_{GWF}$ is then minimized with respect to $\{g(\Gamma_i)\}$ with conjugate gradient method after the analytical $\frac{\partial E_{GWF}}{\partial g(\Gamma_i)}$ is evaluated. The calculation of the gradient is readily partitioned with regard to the configuration $\Gamma_i$, as the evaluation of $\frac{\partial E_{GWF}}{\partial g(\Gamma_i)}$ for a particular $\Gamma_i$ does not involve the evaluation of other derivatives. Therefore, the computational workload of the approach can be easily handled by efficient parallel computing.

For a diatomic molecule, the expressions in Eqs. (6)-(10) are rigorous without any approximations. Next, we generalize our approach for periodic bulk systems (molecules with more than 2 atoms can be treated in the same way). The Hamiltonian for a bulk and periodic system in the second quantization form can be expressed as,

$$H = \sum_{Ii\alpha,Jj\beta,\sigma} t_{Ii\alpha Jj\beta} c^\dagger_{Ii\alpha\sigma} c_{Jj\beta\sigma} + \frac{1}{2} \sum_{\substack{Ii\alpha,Jj\beta \\ Kk\gamma,Ll\delta,\sigma\sigma'}} u(Ii\alpha,Jj\beta;Kk\gamma,Ll\delta) c^\dagger_{Ii\alpha\sigma} c^\dagger_{Jj\beta\sigma'} c_{Ll\delta\sigma'} c_{Kk\gamma\sigma},$$

(11)

where $I,J,K,L$ represent the unit cell indices, $i,j,k,l$ are the atomic site indices, $\alpha,\beta,\gamma,\delta$ the orbital indices, and $\sigma,\sigma'$ the spin indices. The one-electron hopping integral, $t$, and the two-electron Coulomb integral, $u$, are defined similarly to Eq. (2)(3). The GWF has the form,

$$|\Psi_{GWF}\rangle = \prod_{Ii} \left( \sum_\Gamma g(\Gamma_{Ii}) |\Gamma_{Ii}\rangle \langle \Gamma_{Ii} | \right) |\Psi_0\rangle,$$

(12)

where $g(\Gamma_{Ii})$ is the Gutzwiller variational parameter determining the occupation probability of the on-site configuration $|\Gamma_{Ii}\rangle$, defined as a Fock state at the $i^{th}$ site of the $I$ th unit cell: $|\Gamma_{Ii}\rangle \equiv \prod_{\alpha\sigma \in \Gamma_{Ii}} c^\dagger_{\alpha\sigma} |\varnothing\rangle$. Here the creation operator $c^\dagger_{\alpha\sigma}$ creates an electron at the orbital-$\alpha$ with spin-$\sigma$ in the vacuum state $|\varnothing\rangle$. Since all unit cells are identical, $g(\Gamma_{Ii})$ does not depend on a specific unit cell and can be written as $g(\Gamma_{Ii}) = g(\Gamma_i)$. The total energy can be expressed as,

$$\begin{aligned} E_{GWF} &= \sum_{Ii\alpha,Jj\beta,\sigma} t_{Ii\alpha,Jj\beta} \langle c^\dagger_{Ii\alpha\sigma} c_{Jj\beta\sigma} \rangle_{GWF} + \frac{1}{2} \sum_{\substack{Ii\alpha,Jj\beta \\ Kk\gamma,Ll\delta,\sigma\sigma'}} u(Ii\alpha,Jj\beta;Kk\gamma,Ll\delta) \langle c^\dagger_{Ii\alpha\sigma} c^\dagger_{Jj\beta\sigma'} c_{Ll\delta\sigma'} c_{Kk\gamma\sigma} \rangle_{GWF} \\ &\approx \sum_{Ii\alpha,Jj\beta,\sigma} t_{Ii\alpha,Jj\beta} \langle c^\dagger_{Ii\alpha\sigma} c_{Jj\beta\sigma} \rangle_{GWF} + \frac{1}{2} \sum_{\substack{Ii,\alpha\beta\gamma\delta \\ \sigma\sigma'}} u(Ii\alpha,Ii\beta;Ii\gamma,Ii\delta) \langle c^\dagger_{Ii\alpha\sigma} c^\dagger_{Ii\beta\sigma'} c_{Ii\delta\sigma'} c_{Ii\gamma\sigma} \rangle_{GWF} \\ &+ \frac{1}{2} \sum_{\substack{Ii\alpha,Jj\beta \\ Kk\gamma,Ll\delta,\sigma\sigma'}}{}' \left( u(Ii\alpha,Jj\beta;Kk\gamma,Ll\delta) - \delta_{\sigma\sigma'} u(Ii\alpha,Jj\beta;Ll\delta,Kk\gamma) \right) \langle c^\dagger_{Ii\alpha\sigma} c_{Kk\gamma\sigma} \rangle_{GWF} \langle c^\dagger_{Jj\beta\sigma'} c_{Ll\delta\sigma'} \rangle_{GWF} \end{aligned}$$

(13)

Similar to Eq. (5), the on-site 2PCM are treated rigorously here and the inter-site 2PCM are evaluated using HF-type factorized approximation. In the RESULTS section of this work, however, we will focus on Hubbard models only, where the inter-site Coulomb terms are not included in the Hamiltonian. In this specific case, the HF-type factorized approximation is not needed.

The one-particle density matrix (1PDM) can be expressed as,[3]

$$\left\langle c^\dagger_{Ii\alpha\sigma} c_{Ii\beta\sigma} \right\rangle_{GWF} = \frac{1}{\left\langle \Psi_{GWF} | \Psi_{GWF} \right\rangle_{Ii,Jj}} \sum_{\Gamma_{Ii},\Gamma'_{Ii},\Gamma_{Jj}} \left\langle \Gamma_{Ii} | c^\dagger_{Ii\alpha\sigma} c_{Ii\beta\sigma} | \Gamma'_{Ii} \right\rangle \cdot g(\Gamma_i) g(\Gamma'_i) g(\Gamma_j)^2 \xi_{\Gamma_{Ii},\Gamma_{Jj},\Gamma'_{Ii},\Gamma_{Jj}},$$ (14)

$$\left\langle c^\dagger_{Ii\alpha\sigma} c_{Jj\beta\sigma} \right\rangle_{GWF} = \frac{1}{\left\langle \Psi_{GWF} | \Psi_{GWF} \right\rangle_{Ii,Jj}} \sum_{\Gamma_{Ii},\Gamma_{Jj},\Gamma'_{Ii},\Gamma'_{Jj}} \left\langle \Gamma_{Ii} | c^\dagger_{Ii\alpha\sigma} | \Gamma'_{Ii} \right\rangle \left\langle \Gamma_{Jj} | c_{Jj\beta\sigma} | \Gamma'_{Jj} \right\rangle \cdot g(\Gamma_i) g(\Gamma_j) g(\Gamma'_i) g(\Gamma'_j) \xi_{\Gamma_{Ii},\Gamma_{Jj},\Gamma'_{Ii},\Gamma'_{Jj}}, \text{ for } (I,i) \neq (J,j)$$ (15)

where $\xi_{\Gamma_{Ii},\Gamma_{Jj},\Gamma'_{Ii},\Gamma_{Jj}}$ is the coefficient determined from $|\Psi_0\rangle$ and Gutzwiller variational parameters,

$$\xi_{\Gamma_{Ii},\Gamma_{Jj},\Gamma'_{Ii},\Gamma'_{Jj}} = \sum_{\{\Gamma_{Kk}, Kk \neq Ii, Jj\}} \prod_{Kk} g(\Gamma_k)^2 \left\langle \Psi_0 | \Gamma_{Ii}, \Gamma_{Jj}, \{\Gamma_{Kk}\} \right\rangle \left\langle \Gamma'_{Ii}, \Gamma'_{Jj}, \{\Gamma_{Kk}\} | \Psi_0 \right\rangle$$

(16)

and

---

[3] The atomic site $Jj$ is chosen to be the nearest neighbor of site $Ii$ for the calculation of $\left\langle c^\dagger_{Ii\alpha\sigma} c_{Ii\beta\sigma} \right\rangle_{GWF}$. The same treatment is applied when the on-site 2PCM $\left\langle c^\dagger_{Ii\alpha\sigma} c^\dagger_{Ii\beta\sigma} c_{Ii\gamma\sigma} c_{Ii\delta\sigma} \right\rangle_{GWF}$ is calculated.

$$\langle \Psi_{GWF} | \Psi_{GWF} \rangle_{Ii,Jj} = \sum_{\Gamma_{Ii},\Gamma_{Jj}} \xi_{\Gamma_{Ii},\Gamma_{Jj},\Gamma_{Ii},\Gamma_{Jj}} g(\Gamma_i)^2 g(\Gamma_j)^2 . \tag{17}$$

The on-site 2PCM can be expressed as,

$$\langle c_{Ii\alpha\sigma}^{\dagger} c_{Ii\beta\sigma'}^{\dagger} c_{Ii\gamma\sigma'} c_{Ii\delta\sigma} \rangle_{GWF} = \frac{1}{\langle \Psi_{GWF} | \Psi_{GWF} \rangle_{Ii,Jj}} \sum_{\Gamma_{Ii},\Gamma'_{Ii},\Gamma_{Jj}} \langle \Gamma_{Ii} | c_{Ii\alpha\sigma}^{\dagger} c_{Ii\beta\sigma'}^{\dagger} c_{Ii\gamma\sigma'} c_{Ii\delta\sigma} | \Gamma'_{Ii} \rangle g(\Gamma_i) g(\Gamma'_i) g(\Gamma_j)^2 \xi_{\Gamma_{Ii},\Gamma_{Jj},\Gamma'_{Ii},\Gamma_{Jj}}. \tag{18}$$

By comparing Eq. (14)-(18) with Eq. (6)-(10), one can find that the expressions of 1PDM and onsite 2PCM of a bulk system is very close to that of a diatomic molecule, except that the expression of $\xi$ is different. $\xi^0_{\Gamma_i,\Gamma_j,\Gamma'_i,\Gamma'_j}$ as defined in Eq. (8) is now replaced with $\xi_{\Gamma_{Ii},\Gamma_{Jj},\Gamma'_{Ii},\Gamma'_{Jj}}$ in Eq. (16), determined from both $|\Psi_0\rangle$ and the Gutzwiller variational parameters. Clearly, the computational time to rigorously evaluate the expectation values of an operator such as 1PDM (Eq. (14)(15)), the norm of GWF (Eq. (17)), or equivalently the coefficient tensor $\xi$ (Eq. (16)), grows exponentially with respect to the number of atomic sites, as the summation goes through all of them. Therefore, effective approximations to evaluate $\xi_{\Gamma_{Ii},\Gamma_{Jj},\Gamma'_{Ii},\Gamma'_{Jj}}$ has to be adopted. A straightforward way is to approximately evaluate $\xi_{\Gamma_{Ii},\Gamma_{Jj},\Gamma'_{Ii},\Gamma'_{Jj}}$ from $\xi^0_{\Gamma_{Ii},\Gamma_{Jj},\Gamma'_{Ii},\Gamma'_{Jj}}$ defined similarly as in Eq. (8), which can be rewritten as follows,

$$\begin{aligned} \xi^0_{\Gamma_{Ii},\Gamma_{Jj},\Gamma'_{Ii},\Gamma'_{Jj}} &= \langle \Psi_0 | \Gamma_{Ii},\Gamma_{Jj} \rangle \langle \Gamma'_{Ii},\Gamma'_{Jj} | \Psi_0 \rangle \\ &= \sum_{\{\Gamma_{Kk}, Kk \neq Ii,Jj\}} \langle \Psi_0 | \Gamma_{Ii},\Gamma_{Jj},\{\Gamma_{Kk}\} \rangle \langle \Gamma'_{Ii},\Gamma'_{Jj},\{\Gamma_{Kk}\} | \Psi_0 \rangle. \end{aligned} \tag{19}$$

One can see from Eq. (16) that $\xi_{\Gamma_{Ii},\Gamma_{Jj},\Gamma'_{Ii},\Gamma'_{Jj}}$ is a polynomial of $\{g(\Gamma_k)\}$, the degree of which increases with system size. The polynomial will not converge as the system gets larger. However, as $\xi_{\Gamma_{Ii},\Gamma_{Jj},\Gamma'_{Ii},\Gamma'_{Jj}}$ appears in both the numerator and denominator of expression to calculate 1PDM and 2PCM (see Eq. (6)(7)(10)), the result only depends on the relative ratio between the $\xi$s instead of the absolute values of them.

Let $n_{Ii}$ and $n_{Jj}$ be the number of electrons of the configuration $\Gamma_{Ii}$ and $\Gamma_{Jj}$, respectively, and $n'_{Ii}$ and $n'_{Jj}$ be the number of electrons of the configuration $\Gamma'_{Ii}$ and $\Gamma'_{Jj}$, respectively. From the definition of $\xi_{\Gamma_{Ii},\Gamma_{Jj},\Gamma'_{Ii},\Gamma'_{Jj}}$ given by Eq. (11), $n_{Ii}+n_{Jj}=n'_{Ii}+n'_{Jj}$ must hold. We propose that the ratio between the $\xi$s depends on the ratio between the corresponding $\xi^0$s and the number of electrons that occupy site $Ii$ and $Jj$, i.e. $n_{Ii}+n_{Jj}$ or $n'_{Ii}+n'_{Jj}$. One can always set $\xi_{\Gamma_{Ii},\Gamma_{Jj},\Gamma'_{Ii},\Gamma'_{Jj}} = \xi^0_{\Gamma_{Ii},\Gamma_{Jj},\Gamma'_{Ii},\Gamma'_{Jj}}$ with an arbitrary $n_{Ii}+n_{Jj}=n_0$. Then the problem becomes how to evaluate $\xi_{\Gamma_{Ii},\Gamma_{Jj},\Gamma'_{Ii},\Gamma'_{Jj}}$ when $n_{Ii}+n_{Jj} \neq n_0$. Comparing Eq. (16) and (19), one can find that the difference between $\xi$ and $\xi^0$ is the multiplication of the $g$ factors from the rest sites $\prod_{Kk \neq Ii,Jj} g(\Gamma_k)^2$ in the summation. We consider the case that $n_{Ii}+n_{Jj}$ changes from $n_0$ to $n_0-1$, i.e. one extra electron goes to one of the rest sites. Accordingly, one of $g(\Gamma_k)$'s changes to $g(\Gamma_k^+)$, where $\Gamma_k^+$ denotes the new configuration when one electron adds to the configuration $\Gamma_k$. Then the expression of $\xi_{\Gamma_{Ii},\Gamma_{Jj},\Gamma'_{Ii},\Gamma'_{Jj}}$ when $n_i+n_j=n_0-1$ can be written down as,

$$\xi_{\Gamma_{Ii},\Gamma_{Jj},\Gamma'_{Ii},\Gamma'_{Jj}} \approx \xi^0_{\Gamma_{Ii},\Gamma_{Jj},\Gamma'_{Ii},\Gamma'_{Jj}} \cdot \sum_{\Gamma_k} p(\Gamma_k)\frac{g(\Gamma_k^+)^2}{g(\Gamma_k)^2}, \qquad (20)$$

where we define an auxiliary probability, $p(\Gamma_k)$, as the probability of finding the configuration $\Gamma_k$ at site $Kk$. The physical meaning of Eq. (20) is clear: when one extra electron goes to one of the rest sites labeled with $Kk$ ($Kk \neq Ii, Jj$), which is occupied by the configuration $\Gamma_k$ with the probability of $p(\Gamma_k)$, the extra electron will change the configuration from $\Gamma_k$ to $\Gamma_k^+$, and the corresponding ratio change in $\prod_{Kk \neq Ii,Jj} g(\Gamma_k)^2$ is $\dfrac{g(\Gamma_k^+)^2}{g(\Gamma_k)^2}$. The resulting total modification in $\xi_{\Gamma_{Ii},\Gamma_{Jj},\Gamma'_{Ii},\Gamma'_{Jj}}$ is the summation over the respective changes from each possible $\Gamma_k$.

Similarly, the expression of $\xi_{\Gamma_{Ii},\Gamma_{Jj},\Gamma'_{Ii},\Gamma'_{Jj}}$ when $n_{Ii} + n_{Jj} = n_0 + 1$ can be written down as,

$$\xi_{\Gamma_{Ii},\Gamma_{Jj},\Gamma'_{Ii},\Gamma'_{Jj}} \approx \xi^0_{\Gamma_{Ii},\Gamma_{Jj},\Gamma'_{Ii},\Gamma'_{Jj}} \cdot \sum_{\Gamma_k} p(\Gamma_k) \frac{g(\Gamma_k^-)^2}{g(\Gamma_k)^2}, \qquad (21)$$

where $\Gamma_k^-$ is the configuration when one electron is removed from the site with configuration $\Gamma_k$. When $|n_{Ii} + n_{Jj} - n_0| \geq 2$, the treatment in Eq. (20) or (21) is repeated, adding or removing one electron at a time until the desired number of electrons have been added to/removed from the rest sites. The expression of $\xi_{\Gamma_{Ii},\Gamma_{Jj},\Gamma'_{Ii},\Gamma'_{Jj}}$ can be summarized as follows,

$$\xi_{\Gamma_{Ii},\Gamma_{Jj},\Gamma'_{Ii},\Gamma'_{Jj}} \approx \begin{cases} \xi^0_{\Gamma_{Ii},\Gamma_{Jj},\Gamma'_{Ii},\Gamma'_{Jj}} \cdot \left( \sum_{\Gamma_k} p(\Gamma_k) \dfrac{g(\Gamma_k^+)^2}{g(\Gamma_k)^2} \right)^{-(n_{Ii}+n_{Jj}-n_0)} & n_{Ii}+n_{Jj} < n_0 \\ \xi^0_{\Gamma_{Ii},\Gamma_{Jj},\Gamma'_{Ii},\Gamma'_{Jj}} & n_{Ii}+n_{Jj} = n_0 \\ \xi^0_{\Gamma_{Ii},\Gamma_{Jj},\Gamma'_{Ii},\Gamma'_{Jj}} \cdot \left( \sum_{\Gamma_k} p(\Gamma_k) \dfrac{g(\Gamma_k^-)^2}{g(\Gamma_k)^2} \right)^{n_{Ii}+n_{Jj}-n_0} & n_{Ii}+n_{Jj} > n_0 \end{cases} \qquad (22)$$

As one can see from Eq. (22), the approximate form of $\xi_{\Gamma_{Ii},\Gamma_{Jj},\Gamma'_{Ii},\Gamma'_{Jj}}$ is still a function of $g(\Gamma_i)$. $\langle c^\dagger_{Ii\alpha\sigma} c_{Jj\beta\sigma} \rangle_{GWF}$, $\langle c^\dagger_{Ii\alpha\sigma} c^\dagger_{Ii\beta\sigma'} c_{Ii\gamma\sigma'} c_{Ii\delta\sigma} \rangle_{GWF}$, and then $E_{GWF}$ can be expressed explicitly as a function of $\{g(\Gamma_i)\}$. $E_{GWF}$ is then minimized with respect to $\{g(\Gamma_i)\}$ using conjugate gradient method with analytical gradients $\frac{\partial E_{GWF}}{\partial g(\Gamma_i)}$. As Eq. (22) gives an approximate evaluation of $\xi_{\Gamma_{Ii},\Gamma_{Jj},\Gamma'_{Ii},\Gamma'_{Jj}}$, the estimate of energy in our approach is not rigorous. Therefore, the estimated energy is not necessarily an upper bound of the exact ground state energy although the GWF has the form of a variational function. Our method gives a deterministic value of the minimized energy, unlike QMC methods[18,19,20,28] which are based on sampling schemes to get an estimate of a physical quantity bounded with the statistical error.

In this work, we use the diagonal Gutzwiller projector in Fock states. Without additional modifications, it can only capture the atomic limit with density-density type interaction. To switch from Fock state to local atomic eigen-state to construct the Gutzwiller projector is certainly one way forward to capture the full atomic eigen-state. Nevertheless, it will inevitably increase the numerical complexity. Another way to correct the atomic limit is to explicitly add the atomic wave function, such that the ground state wave function is a linear combination of the Gutzwiller wave function and atomic wave function. We have demonstrated the latter approach in oxygen dimer.[15] Which approach is more efficient to correct the atomic limit for solid systems remains to be investigated.

## 3. RESULTS AND DISCUSSION

In this section we test our GCGM approach with 1D, 2D Hubbard model and a 2-

leg Hubbard ladder and compare our results with state-of-the-art numerical methods such as auxiliary-field QMC (AFQMC),[29,30,31] density-matrix renormalization group (DMRG),[32,33] density matrix embedding theory (DMET),[34,35] etc. Figure 1 shows the 3 types of Hubbard model that we will discuss in this section. Hubbard model is the one of the simplest models of correlated electrons, yet it retains the essence of the profound physical phenomena that challenges our understanding of strongly correlated fermions. The single-band Hubbard Hamiltonian can be expressed as,

$$\hat{H} = -\sum_{IiJj,\sigma} t_{IiJj} c^{\dagger}_{Ii\sigma} c_{Jj\sigma} + U \sum_{Ii} c^{\dagger}_{Ii\uparrow} c_{Ii\uparrow} c^{\dagger}_{Ii\downarrow} c_{Ii\downarrow} , \tag{23}$$

where the 1$^{st}$ term describes the hopping of electrons with spin $\sigma$ between site $Ii$ and site $Jj$ and the 2$^{nd}$ term describes an onsite Coulomb interaction $U$ between two electrons occupying the same site $Ii$. These 2 terms compete as the 1$^{st}$ term, the kinetic part, becomes minimal when electrons are moving as fast as possible, while the 2$^{nd}$ term, the onsite $U$ favors electrons being localized at different sites. This competition is essential to the challenging electronic many-body problem and to our understanding of the many intriguing phenomena such as the substantial resistivity changes across the metal-insulator transition and high transition temperatures in superconductors. Hubbard model, as the simplest model to describe this competition, becomes one of the best acknowledged test beds to assess the accuracy of a variety of modern many-body methods.[21,34,36,37,38] In this work, we restrict our study to the repulsive Hubbard model ($U \geq 0$). We also assume that the hopping includes only nearest-neighbor and second nearest-neighbor terms.

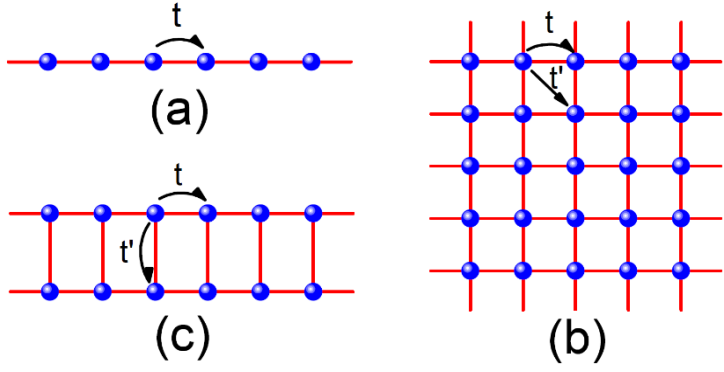

Figure 1. Schematic figures of Hubbard models of (a) 1D chain, (b) 2D square lattice, and (c) 2-leg ladder.

3.1. One dimensional Hubbard model

At each site $i$, there are 4 possible configurations $\Gamma_i = \varnothing, \uparrow, \downarrow \text{ or } \uparrow\downarrow$. For simplicity, we let $g(\varnothing) = g(\uparrow) = g(\downarrow) = 1, g(\uparrow\downarrow) = g$. Then the 1PDM, onsite 2PCM, as well as the total energy, can be readily expressed as a function of $g$ after the probability $p(\Gamma_i)$ is evaluated as a function of $g$. Here, we only need to evaluate $p(\uparrow\downarrow)$, the probability of doubly-occupied site, and the probability of all the other configurations can be deduced from it. Figure 2(a) plots $p(\uparrow\downarrow)$ as a function of $g$. To get $p(\uparrow\downarrow)$, we enumerate all possible configurations $|\Gamma_{11}\Gamma_{21}...\Gamma_{N1}\rangle$ for a 1D ($N =$)18-atom system with scanning $g$ in the range of $0 \leq g \leq 1$ and then find the statistics. We call $p(\uparrow\downarrow)$ obtained in this way "$p_1$". The probability shown in Fig. 2(a) (blue line) is for a half-filling system, while the probability for systems of other fillings can be obtained in a similar way. Next we use $p(\Gamma_i)$ to get $\xi_{\Gamma_{Ii},\Gamma_{Jj},\Gamma'_{Ii},\Gamma'_{Jj}}$ from Eq. (22), where $n_0$ is set to be 2. Finally the total energy can be expressed in terms of $g$ and be minimized. Figure 2(b) plots the ground state energy of 1D standard

Hubbard model, where $t_{ij} = t$ for the nearest neighbor and $0$ otherwise, as a function of $U/t$ for half-filling.

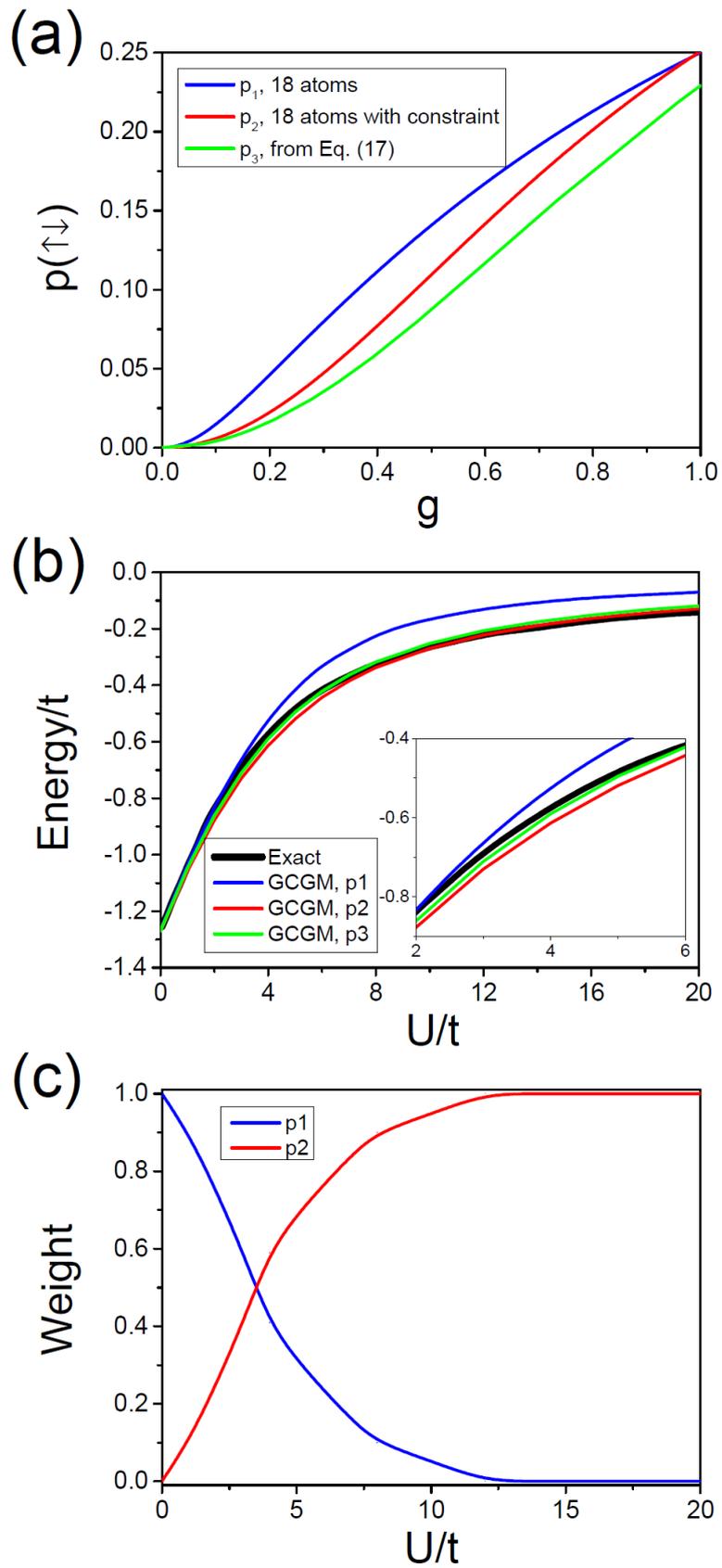

Figure 2. (a) The probability of doubly-occupied site given by 3 methods as a function

of $g$. (b) The ground state energy of 1D standard Hubbard model as a function of $U/t$ at half-filling. The data of the exact Bethe ansatz results is taken from Ref. [21,39]. (c) The weights of the energies calculated from $p_1$ and $p_2$ to reproduce the exact energy.

It can be seen from Fig. 2(b) that the ground state energy calculated from $p_1$ deviates from the exact Bethe ansatz results,[39] especially at the large $U$ region. The error raises from the limitation of GWF: it introduces correlations into the non-interacting wave function only via the on-site correlation factor $g(\Gamma_i)$. However, it overlooks the inter-site correlation. In this case, inter-site correlators like Jastrow correlator[22,28] must be included to achieve satisfying results. As $U/t$ gets large, the hopping of electrons becomes reduced, so doubly-occupied and empty sites tend to be constrained to be adjacent to each other in the ground state, but this correlation is absent in the GWF.[40] To add this correlation, we use a different way to obtain $p(\uparrow\downarrow)$. As before, we enumerate all possible configurations $|\Gamma_{11}\Gamma_{21}...\Gamma_{N1}\rangle$ for a 1D 18-atom system. However, when we enumerate the possible configurations, we only consider configurations with a pairing constraint regulating that each doubly-occupied site must be paired with an empty site, and vice versa. We call $p(\uparrow\downarrow)$ obtained in this way "$p_2$" and plot it in Fig. 2(a) in red line. The ground state energy of 1D Hubbard model as a function of $U/t$ is plotted in Fig. 2(b) in red line. The energy is greatly improved at the large $U$ region, but is somewhat underestimated at the low $U$ region (see the inset of Fig. 2(b)), which makes sense since the pairing constraint is too strong at the weakly correlated regime. If we want to exactly reproduce the energy, we have to "mix" the energies produced by $p_1$ and $p_2$ with the weights as shown in Fig. 2(c). The weights are calculated from the discrepancies between the energies produced by $p_1$/$p_2$ and the exact result. In the weak correlation regime where $U \leq 2$, the $p_1$ energy is dominant indicating that the rigorous GWF is able to describe the weak correlation fairly well. However, when it comes to the strong correlation regime, the $p_2$ energy becomes dominant. This is the regime where we need to add the pairing constraint of doubly-occupied and empty sites in the GWF in order to accurately address the strong inter-site correlation.

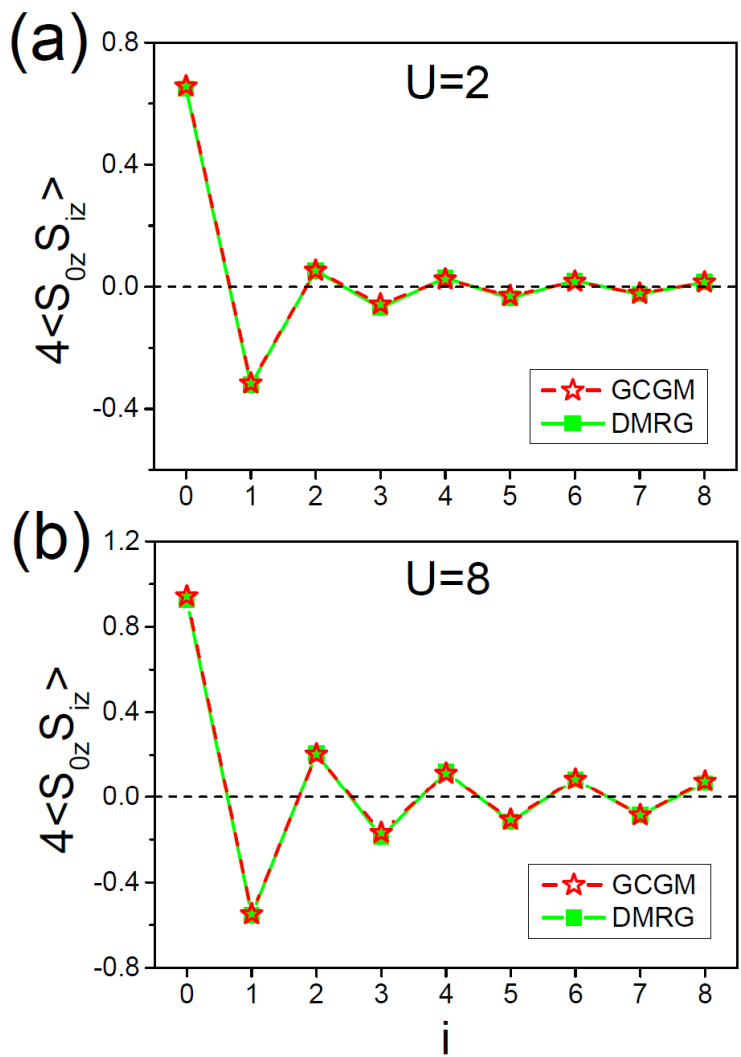

Figure 3. The spin correlation function for (a) $U=2$, and (b) $U=8$ given by the GCGM and the DMRG approach. The DMRG calculations were carried out using the ITensor Library.[41]

Figure 3 plots the spin correlation function in the weak correlation ($U=2$) and strong correlation ($U=8$) regime. The spin correlation function is defined as $4\langle \hat{S}_{0z}\hat{S}_{iz}\rangle = \langle (\hat{n}_{0\uparrow}-\hat{n}_{0\downarrow})(\hat{n}_{i\uparrow}-\hat{n}_{i\downarrow})\rangle$. The DMRG results with 200 atoms are also plotted for comparison. First, the spin correlation is obtained as a function of $g$ in a similar way to obtain $p(\uparrow\downarrow)$ by enumerating all possible configurations for a 1D 18-atom system without and with the pairing constraint. Then the value of $g$ where the energy is minimal is used to get the spin correlation. The same weights without/with the pairing constraint (marked with $p_1/p_2$) as shown in Fig. 2(c) are used to get the combined spin correlation function for the weak and the strong correlation strength.

We propose the 3$^{\text{rd}}$ way to calculate $p(\Gamma_i)$, which can be expressed in terms of $\xi^0_{\Gamma_{Ii},\Gamma_{Jj},\Gamma_{Ii},\Gamma_{Jj}}$ and $g(\Gamma_i)$,

$$p(\Gamma_i) = \frac{\sum_{\Gamma_{Jj}} \xi^0_{\Gamma_{Ii},\Gamma_{Jj},\Gamma_{Ii},\Gamma_{Jj}} g(\Gamma_i)^2 g(\Gamma_j)^2}{\sum_{\Gamma_{Ii},\Gamma_{Jj}} \xi^0_{\Gamma_{Ii},\Gamma_{Jj},\Gamma_{Ii},\Gamma_{Jj}} g(\Gamma_i)^2 g(\Gamma_j)^2}\bigg|_{n_{Ii}+n_{Jj}=2}, \quad (24)$$

where site $Jj$ is site $Ii$'s nearest neighbor. As all unit cells are identical, we can use an arbitrary $I$ in Eq. (24) to estimate $p(\Gamma_i)$. In Eq. (17), when $\Gamma_{Ii}=\uparrow\downarrow$, $\Gamma_{Jj}$ must be $\varnothing$, and vice versa. In other words, when $p(\Gamma_i)$ is evaluated at site $Ii$, a constraint is added requiring that a double occupancy must be paired with a vacancy at site $Ii$ and its nearest neighbor $Jj$, and vice versa. But the constraint is not as strong as the one in "$p_2$", as it does not require the pairing of double occupancy and vacancy at the rest sites. We call the probability evaluated from Eq. (17) "$p_3$". $p_3$ is like a tradeoff between no-constraint at all and the strong pairing constraint for all sites. The energy curve calculated from $p_3$ is plotted in green line in Fig. 2(b). As one can see, $p_3$, in general, provides a good description of $p(\Gamma_i)$. The energy is more accurate at the weak correlation regime than the one calculated from $p_2$, as $p_3$ offers a more reasonable probability for weak correlation. Its performance is not as good as $p_2$, though, at the strong correlation regime where $U$ is bigger than $10$. An advantage of using $p_3$ is that the probability can be directly evaluated from $\xi^0_{\Gamma_{Ii},\Gamma_{Jj},\Gamma_{Ii},\Gamma_{Jj}}$ and $g(\Gamma_i)$ (Eq. (17)), so we do not have to enumerate all possible configurations $|\Gamma_{11}\Gamma_{21}...\Gamma_{N1}\rangle$ to get $p(\Gamma_i)$. It is easy to do the enumeration for 1D Hubbard model, as the trial wave function $|\Psi_0\rangle$ is fixed by symmetry regardless of the correlation strength. However, for more complex systems, $|\Psi_0\rangle$ is dependent on the hopping and Coulomb parameters. It will be a lot of extra work to get the mapping of $p(\Gamma_i)$ to $g(\Gamma_i)$ each time with varying parameters. In the following calculations throughout the paper we will only use $p_3$ as it provides a simple way to determine $p(\Gamma_i)$.

Next we calculate the ground state energy of 1D Hubbard model as a function of $U/t$ with 3 different densities, $n_e = 0.3, 0.7, 1$, where $n_e$ stands for the electron density. For a density that is away from half-filling, we set $n_0$ in Eq. (15) to be the integer that is closest to $2n_e$. For example, when $n_e = 0.3$, $n_0$ is set to be $1$. Let $n_{floor}$ be the largest integer that is no more than $2n_e$, and $n_{ceiling}$ be the smallest integer that is no less than $2n_e$. Eq. (17) can be generalized to,

$$p(\Gamma_i) = w_1 p_{floor}(\Gamma_i) + w_2 p_{ceiling}(\Gamma_i),  \qquad (25)$$

where $w_1 = n_{ceiling} - 2n_e$, $w_2 = 2n_e - n_{floor}$, and

$$p_{floor}(\Gamma_i) = \frac{\sum_{\Gamma_{Jj}} \xi^0_{\Gamma_{Ii},\Gamma_{Jj},\Gamma_{Ii},\Gamma_{Jj}} g(\Gamma_i)^2 g(\Gamma_j)^2}{\sum_{\Gamma_{Ii},\Gamma_{Jj}} \xi^0_{\Gamma_{Ii},\Gamma_{Jj},\Gamma_{Ii},\Gamma_{Jj}} g(\Gamma_i)^2 g(\Gamma_j)^2} \Bigg|_{n_{Ii}+n_{Jj}=n_{floor}}, \qquad (26)$$

$$p_{ceiling}(\Gamma_i) = \frac{\sum_{\Gamma_{Jj}} \xi^0_{\Gamma_{Ii},\Gamma_{Jj},\Gamma_{Ii},\Gamma_{Jj}} g(\Gamma_i)^2 g(\Gamma_j)^2}{\sum_{\Gamma_{Ii},\Gamma_{Jj}} \xi^0_{\Gamma_{Ii},\Gamma_{Jj},\Gamma_{Ii},\Gamma_{Jj}} g(\Gamma_i)^2 g(\Gamma_j)^2} \Bigg|_{n_{Ii}+n_{Jj}=n_{ceiling}}. \qquad (27)$$

The generalized expressions of Eq. (25)-(27) are developed to ensure that the electron density determined from $p(\Gamma_i)$ is $n_e$ when all of $g(\Gamma_i)$ are equal to 1, i.e. recover the electron density at Hartree-Fock limit. The ground state energy as a function of $U/t$ is plotted in Fig. 4(a) with $n_e = 0.3, 0.7, 1$ in comparison with exact results. The results with using GA are also plotted for comparison. The GCGM energy with 3 different fillings are in general in good agreement with exact data, with the exception of the strong correlation regime $U/t > 10$, where GCGM slightly overestimates the energy. The discrepancy comes from the fact, as we discussed in the half-filling case, that the constraint of $p_3$ is not as strong as the pairing constraint that a double occupancy must be paired with a vacancy, which is usually the case at the very strong correlation regime. GA produces the energy curve in reasonable agreement with exact result when the density is $0.3$ and $0.7$. With these densities, GCGM and GA produce very similar energy. However, GA significantly overestimates the energy with half-filling, especially at high correlation regime. The significant discrepancy demonstrates that GA, or the inter-site decoupling type of approximation, is a major source of the error. The GCGM method, by going beyond the Gutzwiller approximation, can achieve an improved accuracy while retaining the computational efficiency of the GWF-based approaches. Fig. 4(b) shows the double occupancy given by GCGM as a function of $U/t$, also in reasonable agreement with exact results.

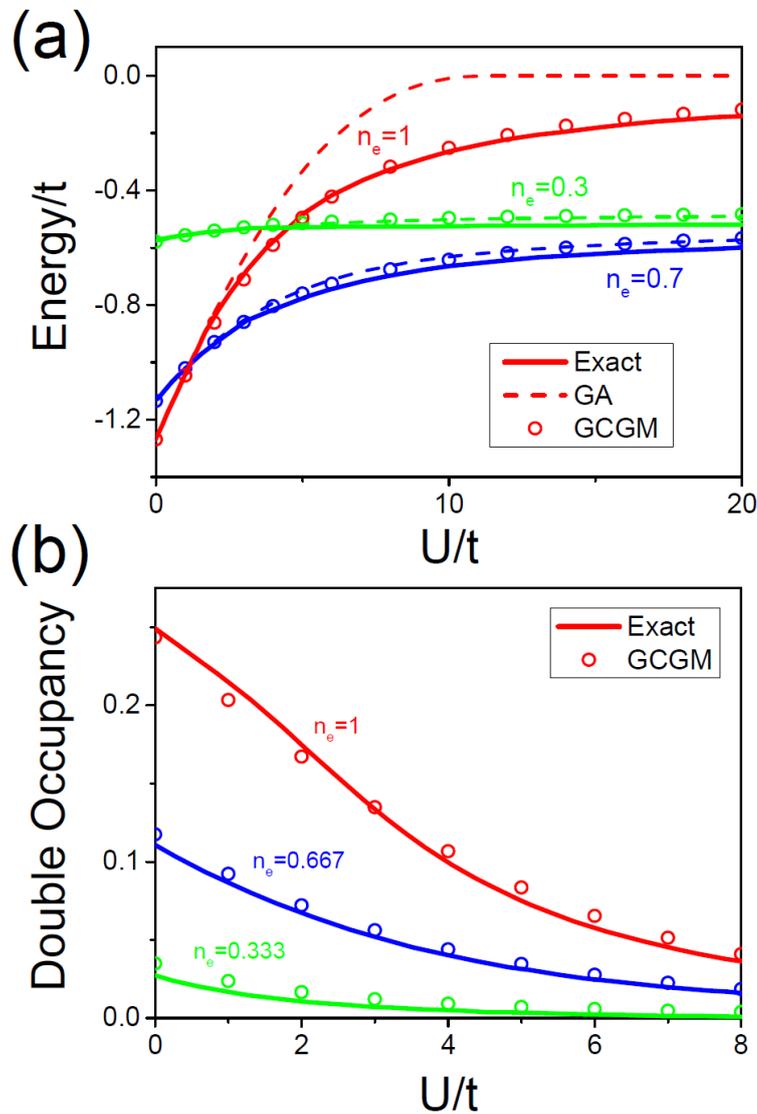

Figure 4. (a) The ground state energy of 1D Hubbard model as a function of $U/t$ with 3 different densities $n_e = 0.3, 0.7, 1$. The data of the exact results is taken from Ref. [21,39]. (b) Double occupancy as a function of $U/t$ with $n_e = \frac{1}{3}, \frac{2}{3}, 1$. The data of the exact results is taken from Ref. [34].

3.2. Two dimensional Hubbard model

We restrict our discussion to the square lattice and to interaction strengths ranging from $U/t = 2$ to $8$. Figure 5(a) plots the ground state energy of a 2D standard Hubbard model as a function of $U/t$ with 3 different densities $n_e = 0.8, 0.875, 1$. For all 3 fillings, the GCGM method yields results in reasonable agreement with the AFQMC, DMRG, and/or DMET results extrapolated to the thermodynamic limit from Ref. [36]. GA, however, fails to yield accurate energy for all 3 densities. The discrepancy is most significant with $n_e = 1$, i.e. the half-filling. Figure 5(b) plots the double occupancy as function of $U/t$ at half-filling in comparison with AFQMC results from Ref. [36] (the results of the other 2 filling are not shown here as we do not have reference data to compare with). Some discrepancy, although not significant, is observed between the GCGM and AFQMC results.

Reference [36] also presented a test case of frustrated $t-t'$ Hubbard model, where $t'$ stands for the 2nd nearest-neighbor hopping ($t_{ij}$ is $t$ for the 1st nearest neighbor, $t'$ for the 2nd nearest neighbor and $0$ otherwise in Eq. (16), also see Fig. 1(b)). Here, we use the same test case to examine the accuracy of the GCGM approach for a frustrated Hubbard model. The parameters are $U/t = 8, n_e = 1, t'/t = -0.2$. The energy obtained by GCGM and GA is summarized in Table 1, as well as the results from some state-of-the-art numerical algorithms, such as AFQMC, DMRG, DMET, unrestricted coupled cluster theory including singles and doubles (UCCSD),[42,43] multireference projected Hartree-Fock (MRPHF)[44,45] and diffusion Monte Carlo based on a fixed-node approximation (FN).[46,47,48] The AFQMC, DMRG, DMET and FN results are extrapolated to the thermodynamic limit. The MRPHF result is for a $8 \times 8$ lattice and UCCSD $10 \times 10$. Again, the GCGM method yields energy in satisfying agreement with other methods, while GA significantly overestimates the energy.

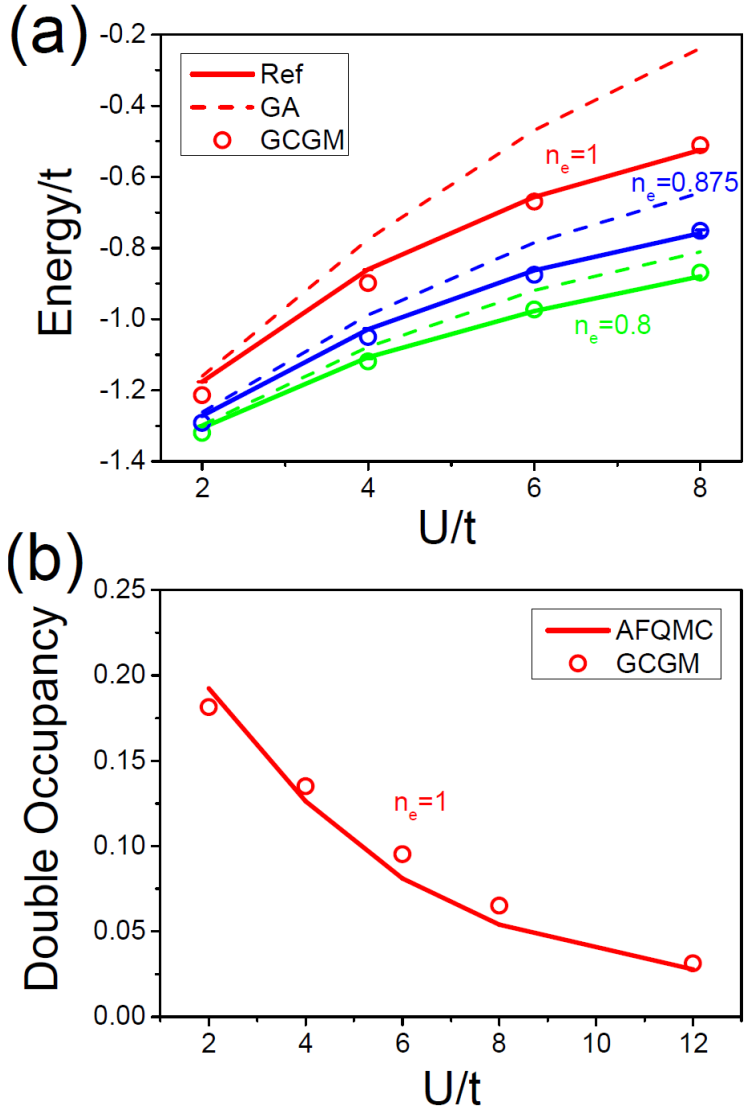

Figure 5. (a) The ground state energy of 2D standard Hubbard model as a function of $U/t$ with 3 different densities $n_e = 0.8, 0.875, 1$. The Reference results are taken from Ref. [36].[4] (b) Double occupancy as a function of $U/t$ at half filling. The AFQMC results are taken from Ref. [36].

| | GCGM | GA | AFQMC | DMET | DMRG | UCCSD | MRPHF | FN |
| --- | --- | --- | --- | --- | --- | --- | --- | --- |

---

[4] As a reference, we use the average energy obtained by AFQMC, DMRG and DMET from Ref. [36]. The AFQMC and DMRG results are missing at some data points, where the DMET energy is just used as the reference.

| | | | | | | | | |
|---|---|---|---|---|---|---|---|---|
| Energy/t | -0.5225 | -0.2507 | -0.5245 | -0.5252 | -0.525 | -0.5199 | -0.518 | -0.5236 |

Table 1. Energy of a 2D frustrated Hubbard model for $U/t=8, n_e=1, t'/t=-0.2$ obtained from GCGM, GA and several many-body methods. Data of AFQMC, DMET, DMRG, UCCSD, MRPHF and FN are taken from Ref. [36].

3.3. Two-leg Hubbard ladder

Now we test our approach with a 2-leg Hubbard ladder as shown in Fig. 1(c), which can be viewed as a special case of a 2-band model. Here, we restrict our discussion to densities $n_e=0.3, 0.7, 1$ and to interaction strengths ranging from $U/t=0$ to $8$. Figure 6 plots the total energy obtained from GCGM and GA as a function of $U/t$. We consider 3 different 2$^{nd}$-nearest-neighbor hopping strength: $t'/t=0.4, 1, 2$. The energy obtained from DMRG with 200 atoms is also plotted as a reference. The DMRG calculations were carried out using the ITensor Library.[41] In general, GCGM yields energy in reasonable agreement with DMRG. Some discrepancy is observed at high $U$ region at $t'/t=2$ with half-filling. The discrepancy between GA and DMRG results is larger. GA also produces energy in reasonable agreement with DMRG with $n_e=0.3, 0.7$, but it significantly overestimates the energy with $n_e=1$. This is consistent with our previous calculations that GA is most inaccurate with half-filling.

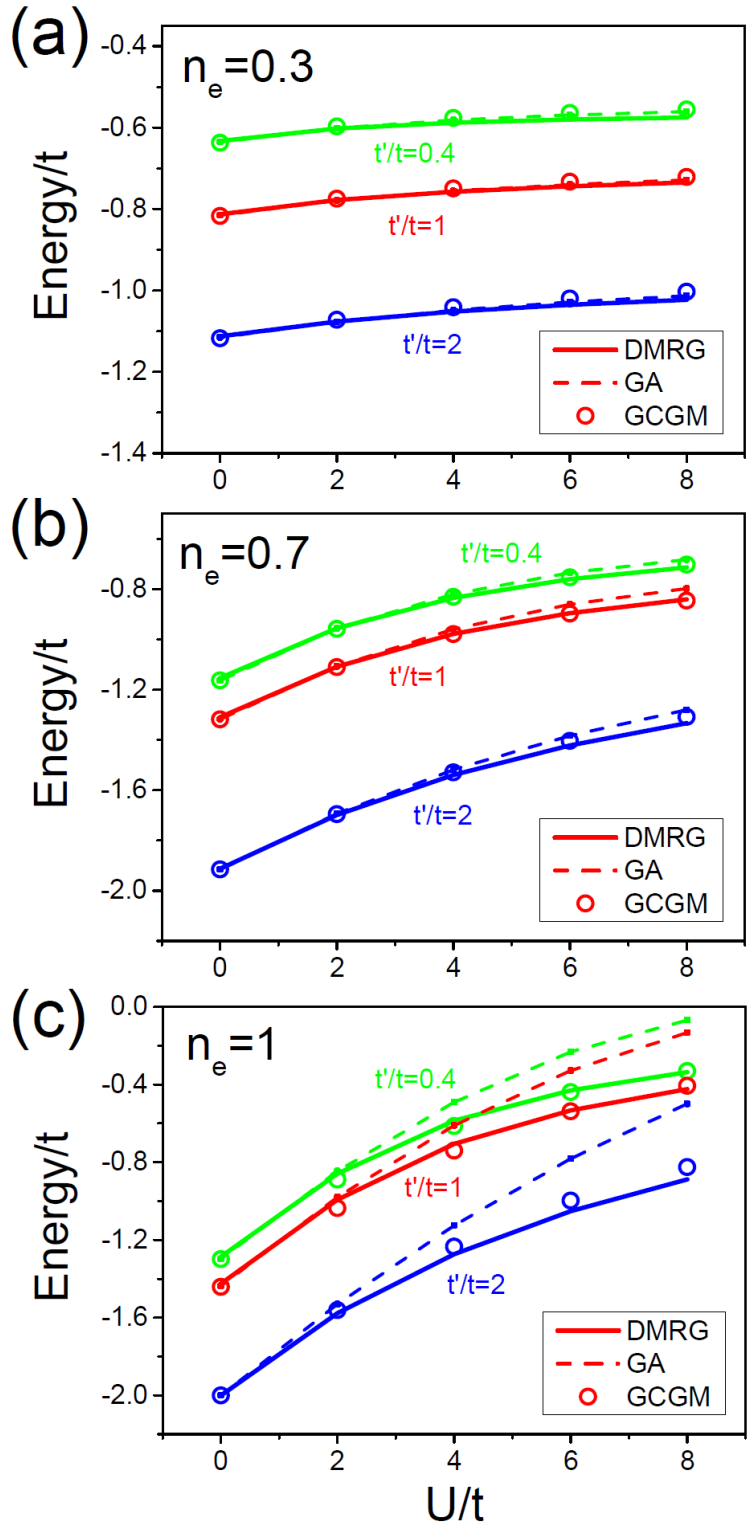

Figure 6. The ground state energy of 2-leg Hubbard ladder as a function of $U/t$ with 3 different densities: (a) $n_e = 0.3$, (b) $n_e = 0.7$, and (c) $n_e = 1$. The DMRG calculations were carried out using the ITensor Library.[41]

3.4. Efficiency of GCGM

Finally, we discuss the computational efficiency and the scaling of the GCGM method. We have demonstrated that the GCGM method can achieve a much improved accuracy by going beyond the Gutzwiller approximation. It is also very efficient. For example, for energy calculations of a 2D Hubbard model, the typical computational time is within a minute, when 1 single core and $19 \times 19$ $k$-points are used.

Then we discuss the scaling of the GCGM method with system size. From Eq. (5), one needs to evaluate the 1PDM $\langle c^\dagger_{Ii\alpha\sigma} c_{Jj\beta\sigma} \rangle_{GWF}$ and on-site 2PCM $\langle c^\dagger_{Ii\alpha\sigma} c^\dagger_{Ii\beta\sigma'} c_{Ii\delta\sigma'} c_{Ii\gamma\sigma} \rangle_{GWF}$ to get the energy. For a periodic bulk system with $n$ atoms per unit cell, if $N$ $k$-points or equivalently $N \cdot n$ atoms are used, the number of on-site 2PCM terms is $N \cdot n$. The number of 1PDM terms would be the number of pairs of atoms to be considered in the system, i.e. also $N \cdot n$. So the GCGM method should scale linearly with system size for periodic bulk systems. In Ref. [15], we tested the scaling of GCGM and we found that GCGM scales a little more than linearly with system size. The extra computational time is attributed to the evaluation of, for example, the predetermined factors $\xi^0_{\Gamma_{Ii}, \Gamma_{Jj}, \Gamma'_{Ii}, \Gamma'_{Jj}}$, which scales more than linearly but needs to be done only once as these factors are constants during the energy minimization.

4. CONCLUSIONS

In this work, we have presented a detailed formalism of the GCGM method for bulk systems and benchmarked it with 1D and 2D single-band Hubbard models at correlation strengths ranging from weak to strong, and at various carrier densities. We have focused on Hubbard models for our benchmark tests for two reasons. First, Hubbard models, both 1D and 2D, are important paradigm models of quantum condensed-matter theory for systems with a strongly screened Coulomb interaction. In spite of having a simple form, they capture the essential feature of strongly correlated systems, i.e. the competition between the kinetic energy that favors electrons being as mobile as possible and the onsite correlation part that favors electrons being localized. Second, they have been widely studied by several state-of-the-art numerical methods and there are plenty of benchmark data in literature that we can compare our results with. In this work, we focus on the observables such as the energy and double occupancy. By comparing the results with those given by GA, we demonstrate that our GCGM method can achieve an improved accuracy through bypassing the commonly adopted GA. The method produces results in satisfying agreement with other pioneering many-body methods. In spite of the fact that GCGM is not as accurate as methods such as QMC or DMRG, it is much faster; it scales only a little more than linearly with system size and is developed under periodic boundary condition with no need of extrapolations to the thermodynamic limit. The further benchmark applications of GCGM in resolving multiple competing phases in model systems[38] and the extension to multi-orbital real materials are natural directions where major efforts are currently being devoted.

Acknowledgments

This work was supported by the US Department of Energy (DOE), Office of Science, Basic Energy Sciences, Materials Science and Engineering Division including a grant of computer time at the National Energy Research Scientific Computing Centre (NERSC) in Berkeley. Ames Laboratory is operated for the US DOE by Iowa State University under Contract No. DE-AC02-07CH11358.